\documentclass[final]{aipproc}

\layoutstyle{6x9}

\usepackage{amssymb}

\newcommand{\be}{\begin{equation}}
\newcommand{\ee}{\end{equation}}
\newcommand{\beq}{\begin{equation}}
\newcommand{\eeq}{\end{equation}}
\newcommand{\ba}{\begin{eqnarray}}
\newcommand{\ea}{\end{eqnarray}}

\def\IZ {\mathbb{Z}}

\def\CC {\mathcal{C}}

\def\CF {\mathcal{F}}

\def\CL {\mathcal{L}}

\def\CN {\mathcal{N}}

\def\CZ {\mathcal{Z}}

\begin{document}

\title{The AdS/CFT Correspondence and Non-perturbative QCD \footnote{This article summarizes both a general overview lecture and a technical talk given by JDE in the XIII Mexican School of Particles and Fields held at San Carlos, M\'exico, October 2008.}}


\author{Jos\'e D. Edelstein}{
  address={Departament of Particle Physics and IGFAE, University of
  Santiago de Compostela \\ E-15782, Santiago de Compostela, Spain\vskip1mm}
  ,altaddress={Centro de Estudios Cient{\'\i}ficos (CECS), Casilla 1469,
  Valdivia, Chile}}

\author{Jonathan P. Shock}{
  address={Departament of Particle Physics and IGFAE, University of
  Santiago de Compostela \\ E-15782, Santiago de Compostela, Spain\vskip1mm}}

\author{Dimitrios Zoakos}{
  address={Departament of Particle Physics and IGFAE, University of
  Santiago de Compostela \\ E-15782, Santiago de Compostela, Spain\vskip1mm}}

\begin{abstract}
We aim at providing an overview on the applications of the AdS/CFT correspondence to non-perturbative aspects of non-Abelian gauge theories, addressed to particle physicists. The finite temperature case, in connection with the physics of the quark-gluon plasma, is emphasized.
\end{abstract}

\maketitle

\section{Introduction}

\begin{figure}[h]
\label{QCDphased}
\includegraphics[height=1.8in]{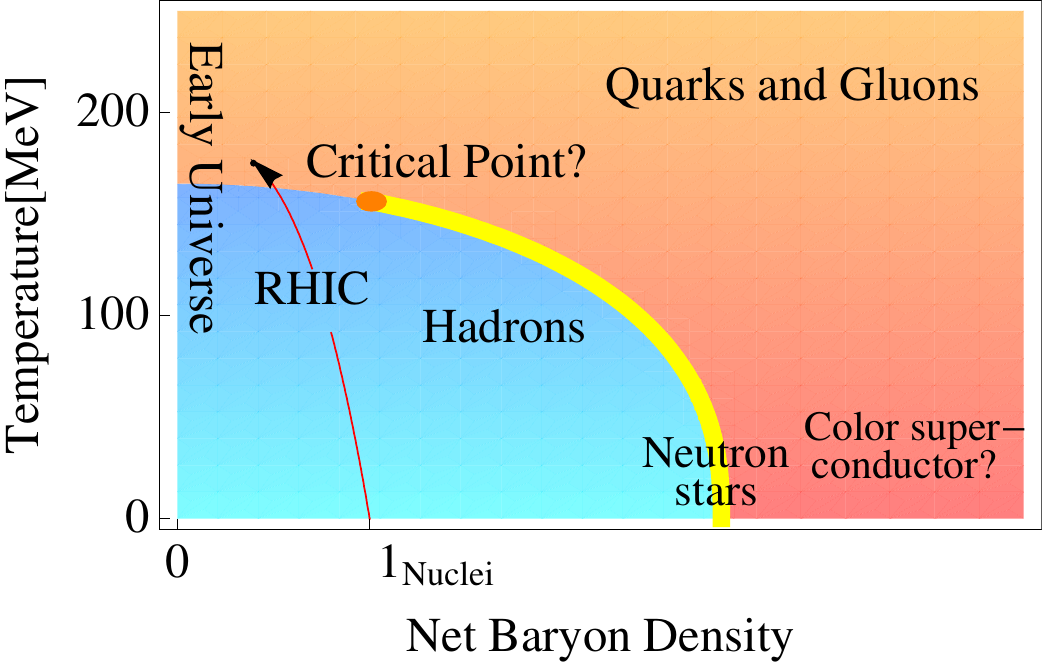}
\caption{The QCD phase diagram. Many corners of the diagram are conjectural.} 
\end{figure}

Many of the hardest problems in non-Abelian gauge theories, such as QCD, are of a non-perturbative nature. We know from the renormalization group equations that whenever a process involves energies of the order of (or lower than) $\Lambda_{\rm QCD}$, the gauge coupling becomes strong enough to spoil the use of perturbative methods. When we move around the QCD phase diagram (figure \ref{QCDphased}), changing the temperature or the baryonic chemical potential, the situation is more complicated and, indeed, not yet fully understood. It is experimentally challenging to explore different regions of this diagram however such programs are currently underway at RHIC and shortly at the LHC.
\vskip1mm
If we focus on the region of vanishing chemical potential, lattice simulations have provided hints of the existence of a critical temperature, T$_c \sim 170$ MeV, where a (rapid) crossover to a deconfined phase occurs (in the large $N_c$ limit, the crossover becomes a first order phase transition). This novel phase, called the quark-gluon plasma (QGP), is believed to have existed during the first few tens of microseconds after the Big Bang. For a long time this was expected to be a weakly coupled regime. However, this expectation was finally contrasted with data once the relativistic heavy ion collider (RHIC) at Brookhaven National Laboratory started to operate a few years ago. Head-on Au+Au collisions at center of mass energies of $\sqrt{s} \simeq 200$ GeV/nucleon, seemingly lead to the production of a tiny drop of QGP. After the collision the {\it fireball} thermalizes very quickly, expands under its own pressure and cools while expanding with a lifetime of order $10$ fm/c.

Indeed it appears that, against the expectations, the QGP is strongly coupled around the critical temperature. Strongly coupled phenomena are generally studied using lattice simulations. However, to date, this approach has not been suitable to apply to real-time dynamical processes such as those involved in this setup. An approach based on a string theoretical construction, the AdS/CFT correspondence, has shown to be a useful tool to scrutinize non-perturbative phenomena of non-Abelian gauge theories, {\it e.g.} those involved in QGP phenomenology. This article aims at reviewing this formalism in a useful manner for particle physicists wanting to be acquainted with what seems to be a powerful and deep tool. After reviewing the main aspects of the AdS/CFT correspondence, and focusing on the case of finite temperature non-Abelian gauge theories, we provide a list of phenomena relevant to the physics of relativistic heavy ion collisions. References to more in-depth explanations are provided where appropriate. Other useful overviews include \cite{Myers2008,Gubser2009}.

\subsection{The Large $N_c$ Limit of Gauge Theories}

The motivation for the AdS/CFT correspondence goes back to an astonishing discovery by 't Hooft \cite{Hooft1974b}, that large $N_c$ gauge theories should have a strongly coupled phase described perturbatively in terms of closed strings. To understand why one might have such a description we consider a simple model with only adjoint fields ({\it e.g.} gluons), using the so-called `double line notation' where the propagators of the fundamental and antifundamental (conjugate) degrees of freedom are written using parallel lines, with arrows indicating the direction of flow of the color degrees of freedom.

With this notation, Feynman diagrams become ribbon graphs, which can be classified topologically in terms of closed Riemann surfaces,  labeled $\Sigma_g$. For this, one must imagine the simplest possible Riemann surface on which the double-line Feynman diagram can be accurately projected. This alone does not give us any new information, but taking the limit such that {$N_c \to \infty$} while keeping the 't Hooft coupling, {$\lambda \equiv g_{YM}^2 N_c$, fixed, we find that these ribbon graphs can be reordered into a series of topological diagrams which will be much simpler than the original Feynman graphs. In these ribbon graphs, vertices (V), propagators  (P) and loops (L) contribute, respectively, with factors of $g_{YM}^{-2}$, $g_{YM}^2$ and $N_c$ to the amplitude of each diagram. Then, the $\lambda$ and $N_c$ contributions to a particular graph can be written in a factorized form as $c_g(\lambda)\, N_c^{\,\chi}$, where $\chi = V - P + L = 2-2g = Eul(\Sigma_g)$ is the Euler character of the given Riemman surface. In figure \ref{PlanarGraphs} we have two diagrams with the same number of vertices and propagators, one of which is planar, and can therefore live on a two dimensional surface without holes, while the other is non-planar and may be drawn on a torus faithfully. The double-line notation makes evident that the latter contributes a factor $N_c^{2}$ less than the former. In the limit that $N_c\rightarrow\infty$ only the planar diagram is important.
\vskip1mm
\begin{figure}[h]
\label{PlanarGraphs}
\includegraphics[height=0.9in]{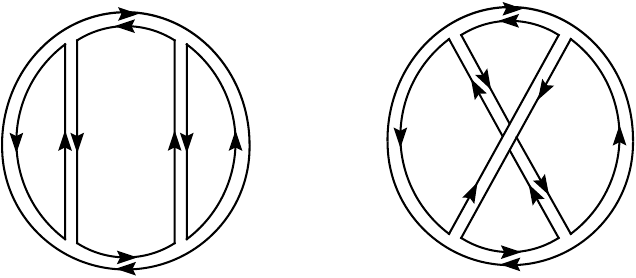}
\caption{Two diagrams which contribute to the same order in the coupling constant, $(g_{YM}^2)^2$, since both have 4 vertices and 6 propagators. The diagram on the left has four loops and will be proportional to $N_c^{\, 2} (g_{YM}^2\, N_c)^2$, whereas the two loops on the right diagram only contribute a factor of $N_c^{\, 0} (g_{YM}^2\, N_c)^2$.} 
\end{figure} 

Any amplitude for gluon interactions, $\mathcal{A}$, can in fact be written as a sum over topologies
\be
\mathcal{A} = \sum_{g=0}^\infty c_g(\lambda) ~N_c^{\, 2 - 2 g}\, ,
\label{sumovert}
\ee
and in the 't Hooft limit only the most trivial topology will survive. It is important to note that while this large $N_c$ gauge theory is clearly not the same as QCD, it has been shown using lattice simulations \cite{Teper2004}, that in many cases the results for large $N_c$ are almost the same as those for $N_c=3$ theories.

We see that there are two regimes for this theory when parameterized by the coupling $\lambda$. For $\lambda \ll 1$ the theory is perturbative in $\lambda$ and we can simply use Feynman diagram summation to calculate amplitudes. However, for $\lambda \gg 1$ an increasing number of diagrams contribute at each order in $N_c$ and the ribbon graphs become dense. They become discretized two dimensional surfaces and we can think of these as the worldsheet of a string theory. Therefore in the 't Hooft limit, not only do we retain only the planar graphs, but also these graphs are described in terms of a simple topology of a string world sheet.
Indeed, the amplitude can be rewritten in string theory notation as
\be
\mathcal{A} = \sum_{g=0}^\infty g_s^{\, 2 g - 2} \mathcal{A}_g (\lambda)\, ,
\qquad\quad g_s \equiv 1/N_c \ll 1 \, ,
\label{sumovtt}
\ee
where $\mathcal{A}_g (\lambda)$ is a {perturbative closed string} amplitude on a Riemman surface $\Sigma_g$, $g_s$ is the string coupling, and $\lambda = g_s N_c$ is a {\em modulus} of the target space. We would like to interpret this as a sum over topologies of the QCD string.
\vskip1mm
\begin{figure}[h]
\label{RangeLambda}
\includegraphics[height=0.55in]{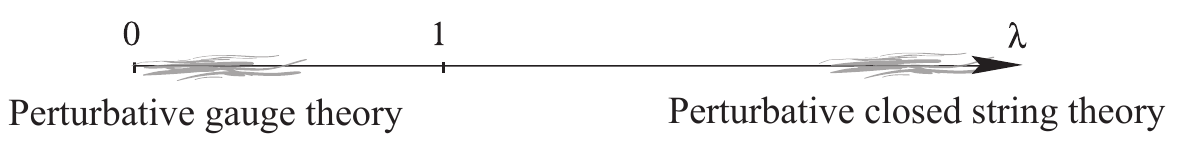}
\caption{Appropriate perturbative description of a planar gauge theory as a function of $\lambda$.} 
\end{figure} 

From the above analysis we can see that the gauge theory admits two complimentary descriptions, as shown in figure \ref{RangeLambda}.
This gives us hints that we may be able to understand QCD in the strong coupling regime using string theory, but clearly we need to know which is the appropriate string theory to describe QCD in this limit. In order to approach this theory we must understand more about explicit constructions of the gauge/string duality.

\subsubsection{Historical Notes}

We sketch here some of the most important points in the history of our understanding of the gauge/string duality in order to comprehend with more clarity how we made such great leaps forward since 't Hooft's original proposal.

In the previous section we saw how a gauge theory at strong coupling could be descibed by a weakly coupled string theory. However, in the early nineties, by looking at the case of lower dimensional matrix models, Polyakov noted that these dual string theories should have an extra dimension \cite{Polyakov1981} on top of those describing the weakly coupled phase. Indeed in the case of QCD in four dimensions, he realized that this may account for the width of the QCD string, while the penetration depth in the fifth dimension would provide a natural scale. Polyakov later noted that the extra dimension had to be warped \cite{Polyakov1998}. On its own, this was not enough to develop a concrete realisation of this duality in four dimensions. However as new ingredients of string theory, called D-branes, were discovered by Polchinski \cite{Polchinski1995} the subject was dramatically altered and the consequences that this could have for such gauge/string dualities were quickly realised. This advance came when Maldacena took Polchinski's D-branes and applied Polyakov's ideas to come up with the first highly non-trivial four-dimensional example of a QFT exactly exhibiting 't Hooft's original ideas of a gauge/string duality \cite{Maldacena1998}.

\section{The AdS/CFT Correspondence}

String theory is not merely a theory of strings. The other ingredients in the theory are now well understood (in some regimes) and include p-dimensional membranes, known as Dp-branes. The possible dimensionality of these objets is fixed by the string theory in question but here we will concentrate on the case of D3-branes, which are 3+1 dimensional hypersurfaces. These are solutions of type IIB string theory, and admit a
two-fold description. 

The first description is in terms of open strings. D3-branes act as hypersurfaces on which open string endpoints can live. One can study the low-energy description of such open strings, which is given by $\CN = 4$ $U(N_c)$ superconformal Yang-Mills (from now on, simply {SYM}) theory. 

Alternatively, from the closed string point of view D3-branes are solitonic solutions of type IIB supergravity. They act as sources for closed strings, whose massless spectrum includes the graviton and the Ramond-Ramond (RR) self-dual five-form field strength. The AdS/CFT correspondence emerges from the complementarity of these open and closed string descriptions of D3-branes. 

\subsection{SYM Theory}

SYM theory in four dimensions (the dimensionality of the world-volume of the D3-branes) has a field content given by a superfield which is composed of one vector, $A_\mu$, six scalars $\phi^I$ ($I=1...6$), and four fermions $\chi_{\alpha}^i$, $\chi_{\dot\alpha}^{\bar i}$ (
$i,\bar i=1,2,3,4$) which are in the ${\bf 4}$ and $\bar {\bf 4}$ representations of the  $SU(4) = SO(6)$ $R$-symmetry group. The SYM Lagrangian contains two free parameters $g_{YM}$ and $\theta_{YM}$,
\be
\CL_{\rm SYM} = Tr \left[ - \frac{1}{2 {g_{YM}^2}} \left( F_{\mu\nu}^2 + \sum_I (D_\mu \phi^I)^2 + \sum_{I,J} [\phi^I,\phi^J]^2 \right) + \frac{\theta_{YM}}{8\pi^2}\, F^{\mu\nu}\, {}^\star F_{\mu\nu} \right] + \CL_\chi\, .
\ee
Not only does SYM not have any explicit scale in the classical Lagrangian, but even quantum mechanically since the theory is superconformal and maximally supersymmetric. The maximal superconformal group in four dimensions is  {$PSU(2,2|4)$} whose bosonic sector is {$SO(4,2) \times SO(6)$}. On top of this large set of global symmetries, the theory displays a strong/weak duality, which acts on the complexified coupling constant as
\be
\tau_{YM} = \frac{\theta_{YM}}{2\pi} + \frac{4\pi i}{g^2_{YM}}\, ,
\qquad\quad \tau_{YM} \to - \frac{1}{\tau_{YM}}\, ,
\ee
which combines with shifts in $\theta_{YM}$ to close the group
{$SL(2,\IZ)$}.

\subsection{Multiple D3-branes in type IIB supergravity}

At low energies where the length of the string is unimportant, type IIB string theory is well approximated by type IIB supergravity. The supergravity solution describing a stack of $N_c$ coincident D3-branes is given by a metric of the form
\begin{equation} \label{D3sol}
ds^2 = \left( 1 + \frac{L^4}{r^4} \right)^{-1/2} \left[ -dt^2 + d\vec x^2
\right] + \left( 1 + \frac{L^4}{r^4} \right)^{1/2} \left[ dr^2 + r^2\,
d\Omega_5^2 \right]\, ,
\end{equation}
plus a constant dilaton $\Phi$, the {\it vev} of which is related to the string coupling ($g_s = e^\Phi$), and $N_c$ units of $F_{[5]}$ flux. $L$ is the only length scale in the solution. This metric interpolates between a throat geometry with curvature dictated by $L$ and an asymptotic ten dimensional Minkowski region.

If we take the {near horizon limit} of the solution given in eq.\eqref{D3sol}, $r \ll L$,
and redefine $z = L^2/r$, we can completely decouple the Minkowski region and are left with a throat geometry which is given by
\be
ds^2 = \frac{L^2}{z^2} \left[ -dt^2 + d\vec x^2 + dz^2 \right] + L^2\,
d\Omega_5^2 \, ,
\ee
which is the Poincar\'e wedge of the direct product of five dimensional anti-de-Sitter space and a five sphere (AdS$_5 \times$ S$^5$). The isometry group of this space is given by $SO(4,2) \times SO(6)$, though if we include fermions, the full supersymmetric isometry group is {$PSU(2,2|4)$}. Note that this is exactly the same as the full global symmetry group of the low energy limit of the open string sector ({\it i.e.} SYM theory). All fields in the closed string sector come in multiplets of this supergroup, while in the open string sector the multiplets define BPS operators \cite{Witten1998}. The   $SL(2,\mathbb{Z})$ weak/strong coupling symmetry of the SYM theory manifests itself when we identify {$g^2_{YM} = 4\pi\, g_s$} and
{$\theta_{YM} = 2\pi \chi$}, where $\chi$  is the vacuum expectation value of the RR scalar (the axion - one of the closed string degrees of freedom which is classically zero for the AdS$_5 \times$ S$^5$ solution). The closed string sector then has precisely the same $SL(2,\mathbb{Z})$ symmetry acting on the dilaton-axion degrees of freedom. 

Because the AdS$_5 \times$ S$^5$ geometry is a BPS solution of the IIB supergravity equations of motion, one can identify the ADM mass of the D3-brane solution with $N_c$ times the tension of a
single D3-brane (the charge), $L^4\, \Omega_5/4\pi\, G_{10} = N_c\, \sqrt{\pi}/\kappa$, where $G_{10}$ is the ten dimensional Newton constant, $\Omega_5 = \pi^3$ is the volume of the 5-sphere, $\kappa = \sqrt{8\pi\, G_{10}} = 8 \pi^{7/2} g_s {\alpha^\prime}^{2}$, and $\alpha^\prime$ is related to the string tension. We see that the radius $L$, of both the AdS throat and the $S^5$, in string units is given in terms of the gauge theory parameters as:
\be
L^4= g^2_{YM}\, N_c\, {\alpha^\prime}^{2}=\lambda{\alpha'}^2 \, .
\ee
Therefore, in order that the stringy modes be unimportant, $L\gg\sqrt{\alpha'}$, which translates into gauge theory language as $\lambda = g^2_{YM}\, N_c \gg 1$. This exactly matches the expectations from 't Hooft's construction and we see that the region where the low energy supergravity limit is valid is exactly where the gauge theory is non-perturbative. This is one the most powerful observations of the gauge/string duality and allows us to understand non-perturbative gauge theories in terms of low energy supergravity. It is important to reiterate here that by analyzing the near horizon (low energy) limit of $N_c$ coincident flat D3-branes, we have accumulated evidence supporting the claim that \cite{Maldacena1998}:

\begin{center}
\fbox{\parbox{13cm}{Planar $SU(N_c)$ SYM theory is dual to type IIB supergravity on AdS$_5 \times$ S$^5$ with $N_c \to \infty$ units of $F_{[5]}$ flux.}}
\end{center}

\subsection{The AdS/CFT Correspondence}

As we have seen in the last section, when one half of the open/closed string dual description of D3-branes is strongly coupled the other is weakly coupled. This means that it is very difficult to prove (or disprove) the conjectured duality. There are indeed three versions of the conjecture. These are:
\begin{itemize}
\item
The strong version: Type IIB string theory on AdS$_5 \times$ S$^5$ ($\forall\, g_s$ and $\forall\, L^2/\alpha^\prime$) is dual to $SU(N_c)$ SYM theory ($\forall\, g_{YM}$ and $\forall N_c$).
\vskip3mm
\item The mild version:
Classical type IIB strings on AdS$_5 \times$ S$^5$ ($g_s \to 0$ and $L^2/\alpha^\prime$ fixed) is dual to planar $SU(N_c)$ SYM theory ($N_c\to\infty$ and $\lambda = g^2_{YM}\, N_c$ fixed).
\vskip3mm
\item The weak version:
Classical type IIB supergravity on AdS$_5 \times$ S$^5$ ($g_s \to 0$ and $L^2/\alpha^\prime \to\infty$) is dual to planar $SU(N_c)$ SYM theory at strong coupling ($N_c\to\infty$ and $\lambda\to\infty)$.
\end{itemize}
Which of these holds is still under debate but more and more evidence is constantly mounting in their favor (most evidence is within the more tractable regime of the weak conjecture).

In order to use the duality to perform calculations, we need a dictionary which relates calculations on one side with calculations on the other. The hope is then that while a calculation may be impossible (or very difficult) on the strongly coupled side, it may be very easy on the weakly coupled, dual side. To establish the {dictionary} between the dual descriptions is an ongoing and very difficult process. We will now comment on some of its most important entries.

\subsection{Correlation Functions}

Here we discuss the dictionary entry which relates correlation functions on either side of the duality. Consider a supergravity field $\Phi_i$ of mass $m_i$ which can be shown to have an asymptotic solution close to the AdS boundary
\beq
\Phi_i(z,x^\mu) \sim \varphi_i(x^\mu) ~z^{\, 4 - \Delta_i}\, , \qquad {\rm where} \quad \Delta_i = 2 + \sqrt{4 + m_i^2\, L^2}\, .
\eeq
The {\it golden entry} in the holographic dictionary relates the partition function of the string theory to the generating functional of correlation functions of the SYM theory. In particular a gauge theory operator $\mathcal{O}_i$ is associated with fluctuations of a (dual) supergravity field, $\Phi_i$, through the following relation \cite{Witten1998,Gubser1998b}
\be
{\CZ_{\rm string} \left[ \Phi_i(z,x^\mu)\bigg|_{z=0}
= \varphi_i(x^\mu) \right] = \bigg< \exp \bigg( \int d^4x ~\varphi_i
\mathcal{O}_i \bigg) \bigg>_{\rm SYM}}\, ,
\label{golden}
\ee
where the boundary value for the supergravity field acts as the source for the field theory operator. The key question then is which field can source which operator. Here we exploit the fact that the global symmetries of the two sides of the correspondence match, to find a single supergravity field matching a single operator to give a product which is a scalar under all the global symmetries. In particular, by looking at the dilatation symmetry of the superconformal group we find that the the operator {$\mathcal{O}_i$ must have conformal dimension $\Delta_i$}. Therefore for any gauge invariant {operator} there exists a corresponding string {state} whose mass is related to the scaling dimension of the former and vice-versa.

For most applications, we are in the classical limit ($g_s \to 0$) and it is sufficient to deal with the saddle-point approximation,
\be
\CZ_{\rm string} \left[ \Phi_i \right] \approx \exp
\left( -\Gamma_{\rm sugra}^{({\rm class})}[\Phi_i] \right)\, .
\ee
%

\subsection{Holography: the Radius/Energy duality}

Another extremely important entry in the AdS/CFT dictionary relates the radial direction of the supergravity geometry to the energy scale in the field theory. The AdS$_5$ factor can be thought of as a warped (codimension one) M$^{1,3}$ space,
\be
ds^2 = \frac{L^2}{z^2} \left[ -dt^2 + d\vec x^2 + dz^2 \right]\, .
\ee
Proper times and distances in the bulk and their field theory counterparts are related through the warp factor by 
\be 
E_{\rm SYM} = \frac{L}{z}\, E_{\rm proper} \,\qquad\,  {\rm and} \,\qquad\, \ell_{\rm SYM} = \frac{z}{L}\, \ell_{\rm proper} \, .
\ee
This means that as  $z \to 0$, we are really describing the UV of the gauge theory. Indeed UV divergences in the field theory are related to IR divergences in the gravity theory \cite{Susskind1998}. The radial direction in AdS$_5$ is described as a {holographic coordinate} that,
from the gauge theory point of view, amounts to the {energy scale}. This is probably one of the deepest aspects of the AdS/CFT correspondence whose implications may be, if correct, as revolutionary to the understanding of spacetime dimensionality as the inclusion of time on an equal footing as space in special relativity.

\subsection{Wilson Loops and the $q \bar q$ potential}

In field theory the Wilson loop is an important tool in understanding properties such as confinement. However, for a strongly coupled gauge theory the calculation of the Wilson loop is very difficult. We may ask whether there is a way to calculate this object using the AdS/CFT correspondence and come up with a dual string theory description of the Wilson loop. In the field theory, the Wilson loop is defined as
\be
W[\CC] = \frac{1}{N_c} {\rm Tr ~P}\left[ e^{\, i \oint_\CC A}\right]\, ,
\ee
where the integral over the gauge field is taken over some closed loop $\CC$ and depends on the representation of the gauge group. In the fundamental representation, it can be thought of as a moving $q \bar q$ pair. Choosing $\CC$ to be a rectangle, with the $q \bar q$ pair a distance $L$ apart and propagating for a time $T$, the potential between the quarks, $V_{q\,\bar q}(L)$, can be read off in the limit that $T\to \infty$, via
\be
\langle W[\mathcal{C}] \rangle = A(L)\, e^{- T\, {\rm V}_{q\,\bar q}(L)}\, .
\ee
From the open string description of the D3-branes, a quark has a string ending on it. In order to study a $q\bar{q}$ pair we introduce a W boson which is associated with the breaking of $U(N_c+1) \to U(N_c)\times U(1)$. From the D3-brane point of view this corresponds to separating one brane from the stack of $N_c+1$ branes. In the 't Hooft limit, the appropriate description of the $N_c$ branes is given by their supergravity solution and we are left with a single D3-brane in the AdS$_5 \times$ S$^5$ background. The $q\bar{q}$ pair is now described by a single string starting and ending on the remaining D3-brane (see figure \ref{WilsonLoop}).
\begin{figure}[ht]\label{WilsonLoop}
\includegraphics[height=1.4in]{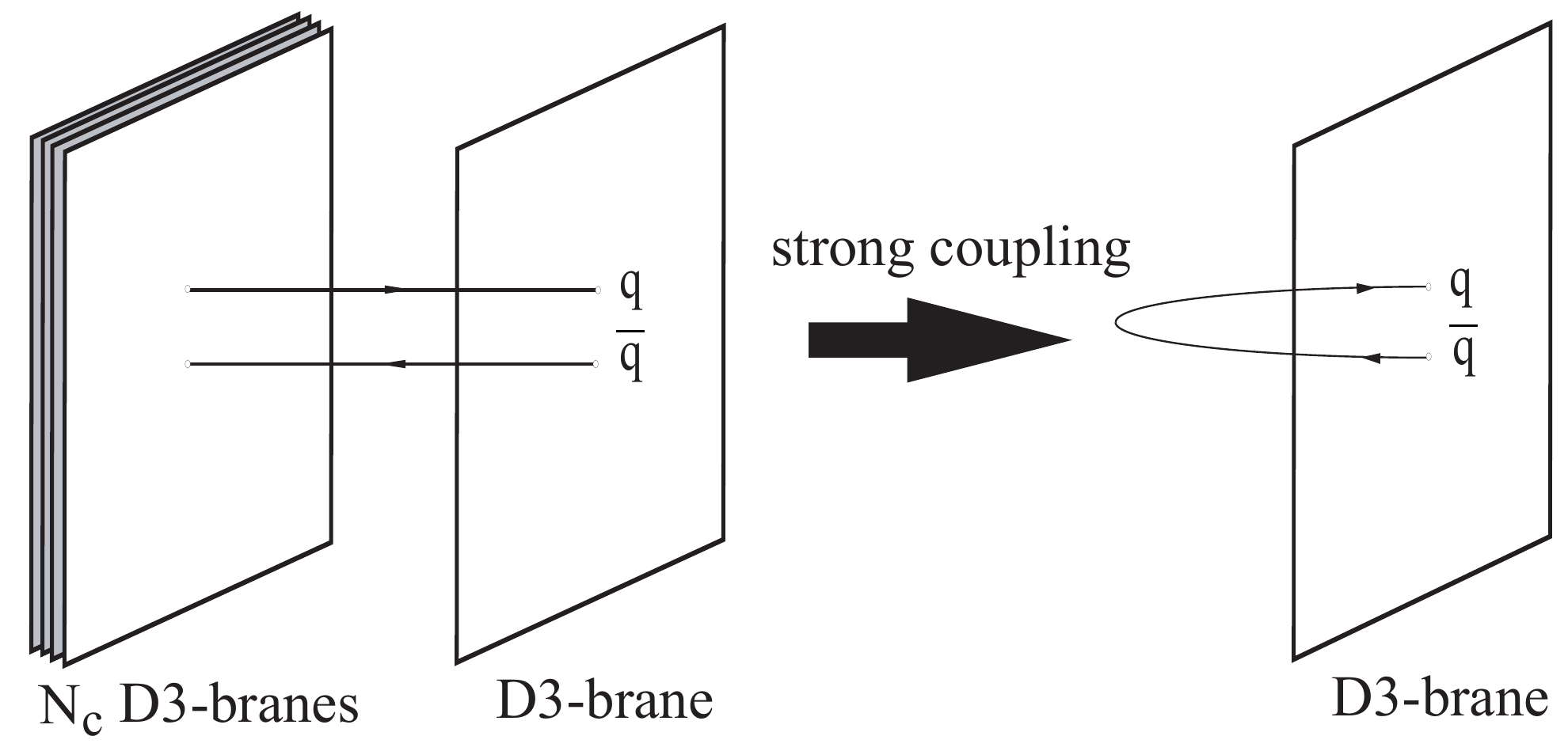}
\caption{We introduce the quark pair in terms
of the W bosons associated to the symmetry breaking $U(N_c+1) \to U(N_c)
\times U(1)$, which we perform by separating a D3--brane. At strong coupling,
the $N_c$ D3--branes are replaced by the background.} 
\end{figure}
Although attached with both ends on the brane, the rest of the string is free to explore the bulk. Given the separation between the ends there is a natural distance that the string probes into the bulk of the geometry, see figure \ref{ConfiningString}. 
\vskip1mm
\begin{figure}[ht] \label{ConfiningString}
\caption{In a theory like QCD, the string explores from the Minkowski boundary to the bulk up to a given radius, this effectively provides a width to the gauge theory confining string.} 
\includegraphics[height=1.1in]{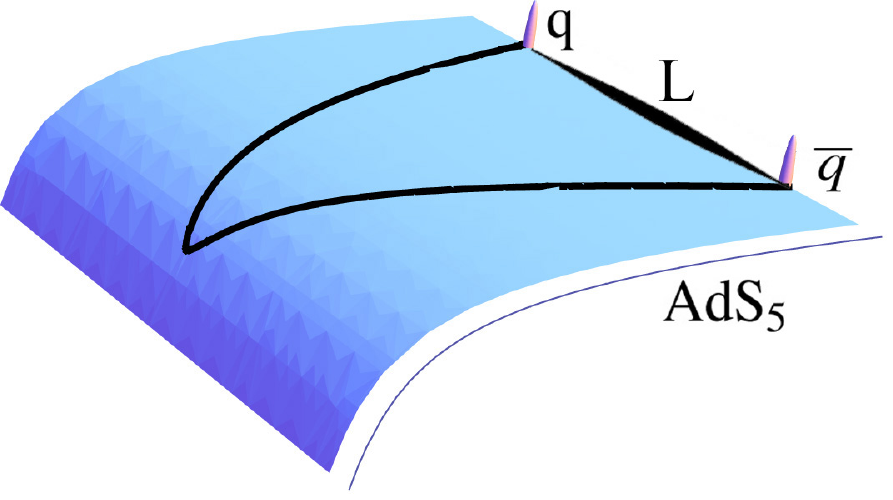}
\end{figure}

The expectation value of the Wilson Loop is then calculated from the partition function of the classical string with boundary conditions specified by $\CC$ taken in the saddle-point approximation. The Nambu-Goto (NG) action gives the solution to the open string catenary problem \cite{Rey2001,Maldacena1998b}
\be
\langle W[\mathcal{C}] \rangle \simeq \exp\,[- \, S_{\rm NG}(\mathcal{C})] \, .
\ee
In the case of SYM theory, we find the very simple result that the $L$ dependence is Coulombic, as expected from the highly restrictive conformal invariance. However, the dependence on the 't Hooft coupling is more subtle, ${\rm V}^{\rm (strong)}_{q\,\bar q}(L) = \CF(\lambda)\,
{\rm V}^{\rm (weak)}_{q\,\bar q}(L)$.
By looking at this potential it is clear that SYM is not a confining theory. As a general rule it is possible to determine whether the field theory dual to a given supergravity background is confining or not, just by minimizing the Nambu-Goto action with given boundary conditions.

\subsection{SYM vs QCD}

Up to now we have been focusing on the supergravity dual of SYM theory described by fields propagating in AdS$_5 \times$ S$^5$. In this case the AdS/CFT dictionary is very well understood and many non-trivial tests have given agreement between the two sides of the correspondence \cite{Klebanov2008a}. Can we however think of SYM as a toy model for QCD? The most important differences between the two theories are:

\begin{itemize}
\item QCD confines while SYM is not confining.
\item QCD has a chiral condensate while SYM has no chiral condensate.
\item QCD has a discrete spectrum while that of SYM is continuous.
\item QCD has a running coupling while SYM has a tunable coupling and is conformal.
\item QCD has quarks while SYM has adjoint matter.
\item QCD is not supersymmetric while SYM is maximally supersymmetric.
\item QCD has $N_c = 3$ while SYM has $N_c \to \infty$.
\end{itemize}

The answer is clearly no! SYM and QCD are two vastly different theories and trying to understand QCD through SYM is not going to get us very far. However, there are ways that we can incrementally change the supergravity theory such that it does describe a better toy model of real QCD.

\subsection{Towards QCD}

It is possible to approach a QCD-like holographic dual, addressing the above issues in a variety of ways. Let us give a flavor of the various possibilities and some helpful references for the interested reader. See \cite{Edelstein2006} for a recent review on this subject.
\begin{itemize}
\item
It is possible to consider multiple D3-branes on curved backgrounds (by replacing the sphere with a more generic manifold), leading to an interesting family of $\CN = 1$ superconformal field theories \cite{Gubser1999,Klebanov1998}. These theories contain matter fields.
\item
The theories above can be further deformed to break conformal symmetry and introduce confinement. This moreover leads to chiral symmetry breaking and a running coupling constant. These theories, in addition, display novel features like a cascade of Seiberg dualities along the renormalization group flow \cite{Klebanov2000}.
\item
Another avenue is to consider higher dimensional D-branes wrapped on certain sub-manifolds of the ten dimensional geometry. It is possible to obtain in this fashion theories that look like $\CN = 1$ supersymmetric Yang-Mills theory in the IR \cite{Maldacena2001,Edelstein2001}.
\item
Various deformations of the original geometry lead to non-supersymmetric, non-conformal gauge theories which exhibit confinement and chiral symmetry breaking \cite{Constable1999a,Polchinski2000,Polchinski2002,Babington2004,Sakai2005}.
\item
Fundamental matter in the quenched approximation, $N_f \ll N_c$, can be added to the gauge theory by introducing D7-branes \cite{Karch2002}, ignoring their effect on the background geometry (see figure \ref{D3D7System}).
\begin{figure}[ht]
\label{D3D7System}
\includegraphics[height=1.5in]{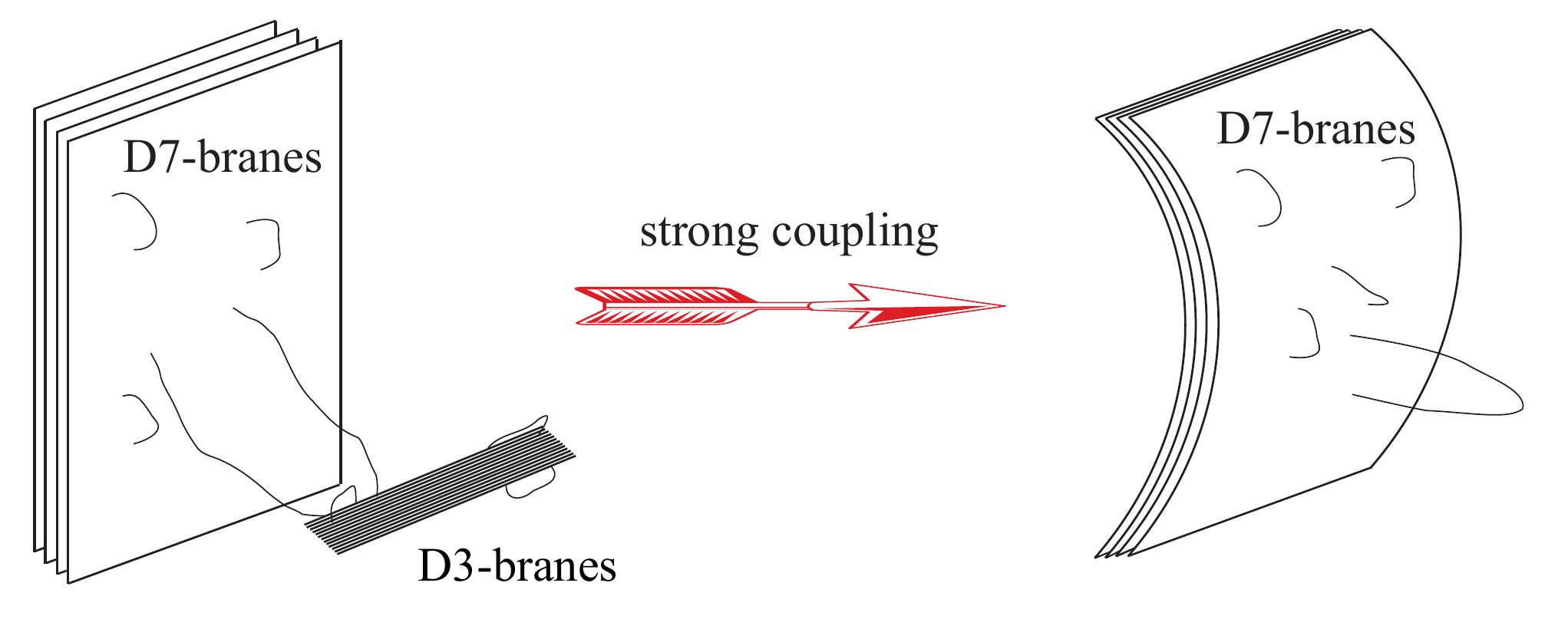}
\caption{A stack of $N_c$ D3-branes and $N_f$ D7-branes. If $N_f \ll N_c$, the near horizon limit replaces the D3-branes by the background AdS$_5$ $\times$ S$^5$, while the D7-branes deform due to the warped geometry. Open strings stretching between a D7 and a D3-brane are in the fundamental of both the color group and the global flavor group, while strings which have both ends on the D7 are singlets under the color group and adjoint under the flavor group. They source the mesons of the SYM theory \cite{Kruczenski2003}.} 
\end{figure}
Adding dynamical quarks, beyond the quenched approximation, is a far harder problem though much progress has also been recently made in this direction \cite{Casero2006}.
\item
Recently, a bottom-up approach has been suggested in which phenomenologically viable models are constructed, motivated by the AdS/CFT machinery but not within the full string theory framework.  The link between cut-off AdS spaces and strongly coupled QCD may be motivated from the light-front holography approach, with interesting results for hadron phenomenology (see \cite{Brodsky2008i} and references therein). This line of research is known as AdS/QCD and has been quite successful in reproducing low energy QCD spectroscopy \cite{Erlich2005,DaRold2005}.
\item
A string theory based approach similar to AdS/QCD uses so-called non-critical constructions in $d\ne 10$ dimensions \cite{Polyakov1999,Klebanov2004}. This makes contact with or attempts to motivate the AdS/QCD approach \cite{Gursoy2008b}.
\end{itemize}

\section{AdS/CFT at Finite Temperature}

\subsection{Thermodynamics}

Introducing temperature $T$ means adding energy without modifying other quantum numbers
such as the brane charges. Gravity duals then involve black branes in an asymptotically AdS background which is simply a black hole embedded in AdS$_5$ \cite{Witten1998}. The ten dimensional metric is given by
\be
ds^2 = \frac{r^2}{L^2} \left[ - \left( 1 - \frac{r_H^4}{r^4} \right) dt^2
+ d\vec x^2 \right] + \frac{L^2}{r^2} \left[ \left( 1 - \frac{r_H^4}{r^4}
\right)^{-1} dr^2 + r^2\, d\Omega_5^2 \right] .
\ee
The asymptotic behavior reveals that the UV physics is not affected by the presence of temperature. However, the IR physics is dramatically modified. The metric above has a
regular event horizon at $r = r_H$ with the Hawking temperature, $T_H$, of the black hole
identified as that of the dual SYM theory, $T = T_H = r_H/\pi\, L^2$. The Bekenstein-Hawking (BH) entropy density of the black hole is proportional to the area of the event horizon \cite{Bekenstein1973,Hawking1974}
\be
s_{\rm BH} = \frac{A_h}{4\hbar\, G} \qquad \Rightarrow \qquad
s_{\rm BH} = \frac{\pi^2}{2}\, N_c^2\, T^3 .
\ee
Notice that the entropy, besides being proportional to $N_c^2$ (a signal of deconfinement), is $3/4$ of the Stefan-Boltzmann value for a free gas of quarks and gluons. This factor is not a mistake of the AdS/CFT calculation, rather it is a prediction for a finite temperature strongly coupled SYM theory. It appears that
\be
s_{\rm SYM} = \CF(\lambda)\, s_{\rm SB}\, , \qquad\quad
\CF(0) = 1\, , \qquad \CF(\infty) = \frac{3}{4}\, .
\ee
Indeed, perturbative computations \cite{Fotopoulos1999,Vazquez-Mozo1999} and string theory corrections (at large $\lambda$) \cite{Gubser1998a}, are consistent with a monotonic function $\CF(\lambda)$.

Lattice calculations of the energy density of QCD displays a plateau with $\epsilon \simeq 4/5\, \epsilon_{\rm SB}$ \cite{Braun-Munzinger2007}. Notice that $4/5$ is closer to $3/4$ than to $1$. The result for SYM loosely suggests that the value of the entropy density in lattice simulations might be interpreted as a fingerprint of a strongly coupled quark-gluon plasma. The question now is whether or not SYM at finite temperature is closer to real QCD than its zero temperature counterpart.

\subsection{SYM vs QCD at Finite Temperature}

The AdS/CFT dictionary is of great help in understanding SYM theory at finite temperature. Just as we did for zero temperature, we can make a comparison between SYM and QCD in the heated phase:
\begin{itemize}
\item None of them is confining, for $T > T_c$.
\item None of them is conformal.
\item QCD is strongly coupled at $T_c < T < 2 T_c$, SYM has a tunable coupling.
\item Both theories display Debye screening.
\item Both theories are non supersymmetric.
\end{itemize}
The finite number of colors and the inclusion of matter in the fundamental representation are clearly not altered by the finite temperature set up and these should be separately addressed as before.

We can see clearly that while the theories are still not identical, there are much stronger similarities than in the zero temperature case and we may hope to gain some more insight into real world finite temperature QCD using the AdS/CFT correspondence than we could do at zero temperature.

\section{AdS/CFT and the Quark-Gluon Plasma}

As discussed in the introduction, the experiment at RHIC is an excellent testing ground which we can use to compare real-world QCD with the results from the AdS/CFT correspondence at finite temperature. There are many properties of the strongly coupled plasma which can be studied but we will concentrate on just a few here, leaving a short discussion of several others for the end.

\subsection{Elliptic Flow}

At heavy ion colliders, in off-center collisions, (impact parameter $b \neq 0$), the heated overlap region is almond shaped. Either free streaming washes out the `almond', or else collective interactions result in an anisotropy of detected particles with respect to the interaction plane. From such collisions (the distribution of which is well understood) we can ask what the number of particles detected at a given angle will be. This is parameterized by
\be
\frac{dN}{d\phi} \approx 1 + {\rm v}_2(A,b,p_t)\, \cos\, 2\phi\, ,
\ee
where the parameter v$_2$, the second moment, which depends on the atomic number of the colliding ions, $A$, the impact parameter, and the transverse momenta of the particles, $p_t$, tells us about the elliptic flow within the medium after the collision. Hydrodynamics seemingly provides a good description of the QGP \cite{Bjorken1983}. The experimental results for v$_2$ can be compared with models which are parameterized as a function of $\eta/s$, the shear viscosity to entropy ratio. The results of such a comparison are sketched in figure \ref{EllipticFlow}.
\begin{figure}[h] \label{EllipticFlow}
\includegraphics[height=1.5in]{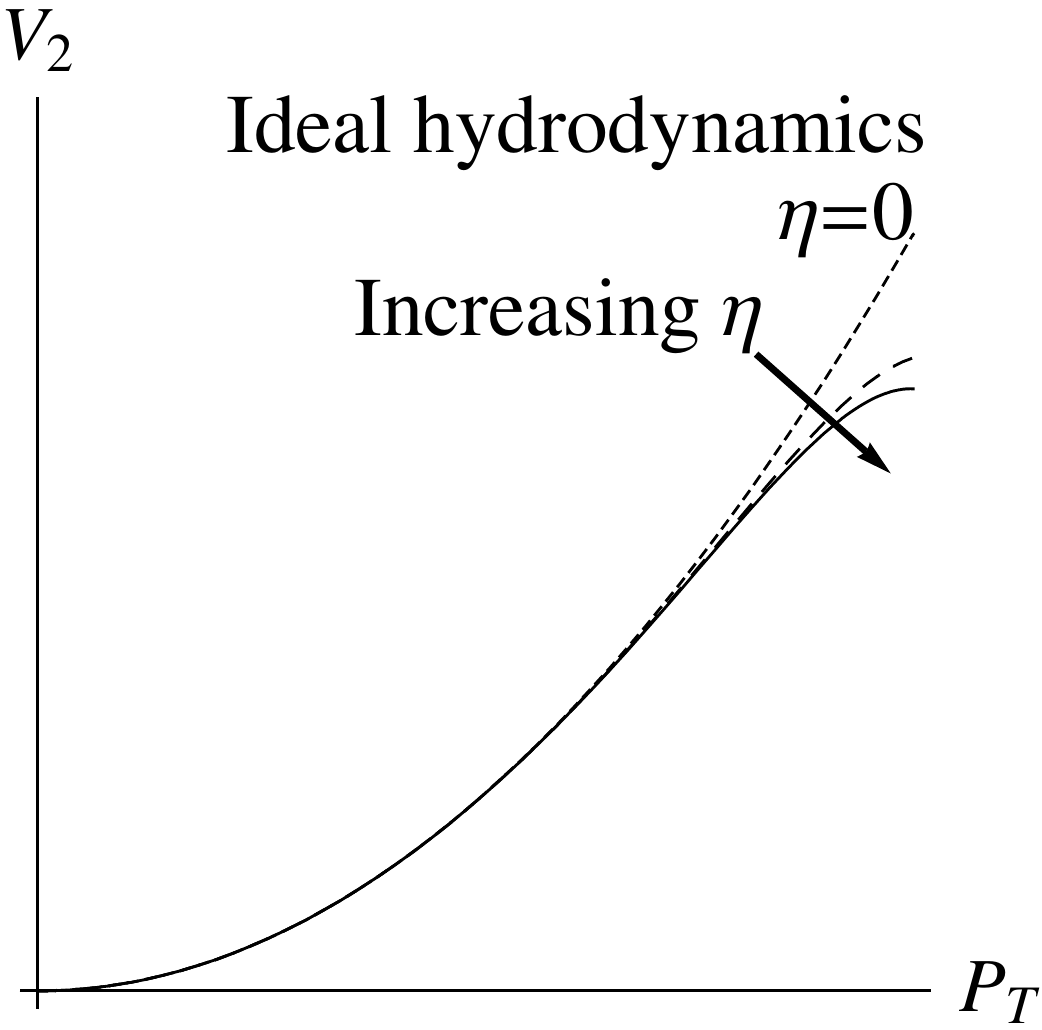}
\caption{Behavior of v$_2$ as a function of $p_t$ ($A$ and $b$ are kept fixed) for varying, small values of $\eta/s$. Hydrodynamics is destroyed for high values of $b$, {\it i.e.} highly off-center collisions.} 
\end{figure}
It is found that the best fit to the data is obtained for $\eta/s=0.15\, \hbar/k_B$ and is certainly below $5/(4 \pi)\, \hbar/k_B$ \cite{Song2008}. The perturbative value of $\eta/s$ goes to $\infty$ for $\lambda \to 0$ \cite{Huot2007}, which is manifestly in conflict with RHIC data, therefore it would be useful to scrutinize its value in the strongly coupled regime. 

The calculation of the shear viscosity comes from studying the finite temperature stress energy tensor which, in the long wavelength limit, is of the form
\be
T_{ij} = p\, \delta_{ij} -
\eta\, \left( \partial_i u_j + \partial_j u_i -{2\over 3} \delta_{ij}\, \partial_k u_k \right)\, ,
\ee
where $p$ is the pressure and $u_i$ is the local fluid velocity. Using the Kubo formula it can be shown that \cite{Policastro2001a}
\be
\eta = \lim_{\omega\to 0} {1\over 2\omega}
\int dt\, d^3 x\; e^{i\omega t} \langle [T_{xy}(t,\vec x), T_{xy}(0,0)] \rangle\, ,
\ee
an expression that can readily be put through the AdS/CFT machinery. Recall that the {\it golden entry} of the AdS/CFT correspondence, eq.\eqref{golden}, allows us to compute just such a two point function. In fact, the calculation amounts to finding the absorption cross section, $\sigma_{\rm abs}$, of a low energy graviton $h_{xy}$ (the field which couples to the stress energy tensor on the boundary) by the black hole, $\eta = \sigma_{\rm abs}(\omega\to 0)/16\pi\, G$ \cite{Gubser1997a,Gubser1997}. Under very general conditions, $\sigma_{\rm abs}(\omega\to 0) = A_H$ \cite{Das1997,Emparan1997}. This, in turn, is proportional to the entropy density. Then \cite{Policastro2001a}
\be
\eta = \frac{\pi}{8}\, N_c^2\, T^3 \qquad \Rightarrow \qquad
{\frac{\eta}{s} = \frac{1}{4\pi}\, \frac{\hbar}{k_B}}\, .
\ee
Based on {universal properties of black hole horizons}, this result appears to be extremely robust. All gauge/gravity duals fulfill this relation, regardless of their field content and number of supersymmetries.

It is tempting and natural, then, to conjecture that it may be valid for QCD (up to finite $\lambda$ and $N_c$ corrections). Going beyond supergravity by including stringy corrections both to $\eta$ and $s$, is consistent with a monotonic interpolation between the perturbative and non-perturbative results \cite{Gubser1998a,Buchel2005}. This is one of the most important, and perhaps surprising results of the attempt to formulate a finite temperature holographic dual of QCD. It is, indeed, strikingly close to the experimental value and is one of the main successes of this approach to the physics of the QGP.

\subsection{Jet Suppression}

Another interesting signature of the QGP that one may study from the AdS/CFT perspective is that of jet suppression. If we consider the ratio, $R$, of the number of jets in heavy ion collisions ({\it e.g.} Au+Au) to that seen in p+p collisions  (scaled to account for the number of participating nucleons) we see that any departure from $R=1$ indicates that partons produced in hard scattering events are slowed by the hot medium which is only present in the Au+Au collisions. Indeed such suppression caused by this medium induced slowdown is detected at RHIC.

The main channel for this energy loss is medium induced gluon radiation (figure \ref{gluonrad}). 
\begin{figure}[h]\label{gluonrad}
\includegraphics[height=1.6in]{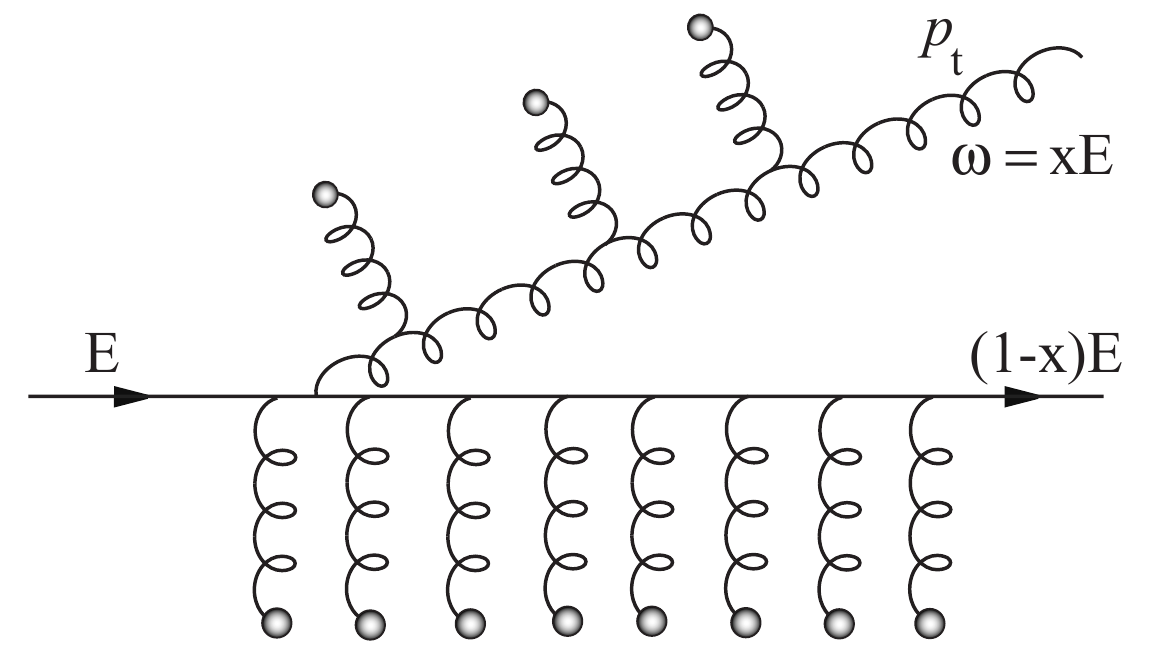}
\caption{Medium induced gluon radiation. A quark with energy $E$ emits a gluon with a fraction of momentum $x$ and transverse momentum $p_t$. In the eikonal approximation, $E\gg xE\gg p_t$.} 
\end{figure}
As it travels through the medium, in the eikonal approximation the particle trajectories can be written as Wilson lines in light-cone coordinates \cite{Baier1998}. In the multiple soft scattering approximation, the transverse position of the gluon follows a Brownian motion, and the {average transverse momentum} after traveling a distance $L$ is characterized by the transport coefficient, ${\hat q} \simeq \langle p^2_t\rangle/L$, called the {\it jet quenching parameter} (see \cite{Casalderrey-Solana2007,Edelstein2008} for a thorough explanation).

Calculating the cross-section for gluon emission \cite{Wiedemann2000} one finds that up to logarithmic corrections in the small distance limit
\be
\frac{1}{N_c^2-1}\, {\rm Tr}\langle W^A({\bf x})\, W^A({\bf y})
\rangle \simeq \exp\left[-\frac{({\bf x-y})^2}{4\sqrt{2}} \int
dx_-\, \hat q(x_-)\right]\, .
\ee
The expectation value of the two point function for the adjoint Wilson loop and the corresponding medium properties are therefore all encoded in this single parameter, $\hat q$. This expression can be extrapolated to the strong coupling limit and considered as a non-perturbative definition of $\hat q$. Written in this limit we have the following definition in terms of the expectation value of the Wilson loop defined on a closed path $\CC$ with a large light like side $L^-$ and a much smaller space-like separation $L\ll L^-$:
\be  \label{jetquench}
\langle W^A(\mathcal{C})\rangle \simeq \exp\left[-\frac{1}{4\sqrt{2}}\,
\hat{q}\, L^-\, L^2 \right]\, .
\label{defqhat}
\ee
As we have seen earlier, we are able to efficiently calculate the expectation value of a Wilson loop using the AdS/CFT machinery. Note that in the large $N_c$ limit we are able to approximate accurately the adjoint Wilson loop by its fundamental counterpart: $\langle W^A(\mathcal{C})\rangle = \langle W^F(\mathcal{C})\rangle^2 + {\cal O} \left( 1/N_c \right)$ and therefore we can use the previous definition for the fundamental Wilson loop in terms of the NG action of the fundamental string.

Performing the Wilson loop calculation using the AdS/CFT prescription, this time in the finite temperature background, we use eq.(\ref{jetquench}) to obtain $\hat q_{{\rm  SYM}} \propto T^3\, \sqrt{\lambda}$ \cite{Liu2006}. In contrast to the weak coupling expression, at strong coupling this does not depend explicitly on the number of degrees of freedom. We may see if this expression agrees with experimental data by considering a representative choice of values for the free parameters:  $\lambda = 6\pi$ ({\it i.e.} $\alpha_{\rm s} = 1/2$) and $T = 300$ MeV. In this case, $\hat q_{SYM} = 4.48$ GeV$^2$/fm, which is reassuringly close to the experimental value $\hat q_{exp} \approx (10 \pm 5)$ GeV$^2$/fm. It is possible to perform the same calculation in different geometries, corresponding to different finite temperature field theories. In contrast to the ratio $\eta/s$, there is no universal value for $\hat q$. Finite 't Hooft coupling corrections tend to reduce $\hat q$, in the direction of the perturbative results, as one would expect for a monotonic function of $\lambda$ \cite{Armesto2006}.

As explained previously, we do not have a string dual of QCD. However, we can try to understand how different quantities depend on supersymmetry, dimensionality, field content, etc. Some attempts to extrapolate results from SYM towards QCD have been explored recently in theories without fundamental degrees of freedom. In order to compare these theories with QCD one should chose an unambiguous quantity to fix in both theories in order to get an accurate comparison. It was suggested that the energy density was an appropriate quantity \cite{Gubser2007}. Choosing this comparison we are led to the following relationship between the SYM and QCD temperatures $T_{\rm SYM} = 3^{-1/4}\, T_{\rm QCD}$ which would mean that $\hat q_{\rm QCD} < \hat q_{\rm SYM}$ \cite{Gubser2008} in contrast to the above first approximation.

It has however been alternatively conjectured that, since the QGP of QCD is approximately conformal at $T \approx 2 T_c$ \cite{Liu2007a},
\be
\frac{\hat q_{\rm QCD}}{\hat q_{\rm SYM}} \simeq
\sqrt{\frac{s_{\rm QCD}}{s_{\rm SYM}}} \approx 0.63\, ,
\ee
which would agree with the above tendency. Clearly the comparison is an area of controversy and the case is still open for discussion. It should also be noted that the addition of flavor in non-critical set-ups tends to challenge the latter result, suggesting that $\hat q_{\rm QCD}$
may be higher than $\hat q_{\rm SYM}$ \cite{Bertoldi2007}.

\section{Further features of the QGP}

There are many subtleties and controversial points that we do not have space to explore in detail here. However we can give some basic ideas of some of the most interesting recent investigations into using the AdS/CFT correspondence to understand several features of the strongly coupled QGP. We provide references to some appropriate literature.

\subsection{Meson melting}

As mentioned, it is possible to introduce fundamental matter into the AdS/CFT correspondence. Doing so allows us to study meson phenomenology and this is particularly interesting at finite temperature. Flavor D7-branes can be embedded into the black D3-brane background in two distinct ways, which give very different effects for the associated mesons \cite{Babington2004,Kirsch2004a}. The D7-brane can either end on the black hole horizon or it can miss the horizon completely. The solutions which do not hit the horizon (corresponding to large values of $m_q/T$) display a discrete spectrum of mesons with a mass gap related to the value of $m_q$. These mesons are completely stable and their spectral function is an infinite sum of delta functions. The solutions that touch the horizon (corresponding to small values of $m_q/T$) on the other hand display a continuous spectral function and the associated mesons have a finite lifetime \cite{Hoyos-Badajoz2007}. By studying this spectrum of quasinormal modes one can calculate the lifetime of the mesons once they enter into the QGP (or, more accurately the SYM plasma as there are no dynamical quarks). In particular it can be seen that while mesons made of light quarks melt quickly into the plasma, heavy quark mesons may remain relatively stable. Indeed it has been noted that $J/\Psi$ mesons do not appear to melt in the RHIC plasma as do the light-quark mesons.

\subsection{Photoproduction}

Photoproduction is an excellent probe of the strongly coupled quark gluon plasma. Once photons are produced, they travel virtually unimpeded through the optically thin medium and can be detected to give an idea of the interior of the hot, dense plasma. Although using the AdS/CFT correspondence we have no gravity dual of a weakly coupled U(1) theory, using the optical theorem we can write the photoproduction rate in terms of a two point function of vector currents, $\bar{q}\gamma_{\mu}q$, which can be calculated by looking at a U(1) vector field on a probe flavor D7-brane \cite{Caron-Huot2006,Mateos2007c,Mas2008a}. Dynamical photon corrections will come at next to leading order to this calculation and therefore we can get a good handle on the true photoproduction rate without a dynamical U(1). The two point function can be written directly in terms of the spectral function calculated from excitations on the D7-brane. The effect of finite baryon density on the photoproduction rate can be calculated and from the spectral function we can study the conductivity and diffusion in the plasma. All of these calculations give more results which can in theory be tested against experiment. One of the major difficulties is that such experiments have a great deal of noise although there are some predicted signals which may be easily seen as sharp peaks in the spectral function \cite{Casalderrey-Solana2008}.

\subsection{Bulk viscosity}

The bulk viscosity is related to the energy loss of a fluid due to its expansion. For a conformally invariant system, $\zeta = 0$. This is the result, for instance, obtained using the AdS/CFT correspondence for SYM. However, as discussed throughout this lecture, there are other gauge/gravity systems where conformal symmetry is not apparent, and they yield a non-vanishing bulk viscosity. An interesting relation among different transport coefficients,
\begin{equation}
\frac{\zeta}{\eta} \gtrsim 2 \left( \frac13 - c_{\rm s}^2 \right) ~,
\end{equation}
where $c_{\rm s}$ is the speed of sound, was obtained within this framework \cite{Buchel2008}. There is a family of gauge theories coming from D-brane configurations that saturate this bound \cite{Mas2007}, a saturation that seems to be a consequence of their kinematical structure \cite{Kanitscheider2009}.

\subsection{Drag force and diffusion wakes}

A high energy particle moving through a hot plasma will lose energy to the surrounding medium, leading to an effective viscous drag on its motion. Within the framework of the AdS/CFT correspondence this is represented by a string hanging from the boundary of AdS into the bulk. The point of attachment carries a fundamental charge under the gauge group $SU(N_c)$, is infinitely massive (when the string endpoint is placed on the boundary) and moves with constant velocity relative to the rest frame of the plasma. The drag force that the trailing string inside the plasma exerts on the external quark is \cite{Herzog2006,Gubser2006,Caceres2006a}
\begin{equation}\label{drag}
\frac{dp}{dt}\,=\,-\frac{\pi \sqrt{g_{YM}^2 N}}{2}\,T^2\,\frac{v}{\sqrt{1-v^2}}\, .
\end{equation}
In contrast to quarks, spinning mesons, {\it i.e.} color singlet states, do not experience any drag effect when propagating through the plasma \cite{Peeters2006,Chernicoff2006}. This means that no force is necessary to keep them moving with fixed velocity. The same applies to baryons \cite{Chernicoff2007}.

Energy transferred to the plasma from the moving quark will cause the energy density of the plasma, in the vicinity of the quark, to deviate from its equilibrium value. That is, the moving quark will create an energy density ``wake'' which moves with it through the plasma. Evaluating the energy density perturbation we observe a net surplus of energy in front of the quark and a net deficit behind. This may naturally be interpreted as plasma being pushed and displaced by the quark, just like the water displacement produced by a moving boat. Another interesting feature revealed by the gravity calculation is the formation of a conical energy wake, or sonic boom, for velocities greater than the speed of sound \cite{Gubser2008a,Chesler2007}. Contrary to quarks where the diffusion wake is strong, when mesons move through the plasma the wake is absent whilst a shock wave is created \cite{Gubser2008c}.

\subsection{Early times after the collision}

One may wonder whether a full description of an ultra-relativistic heavy ion collision may be achieved within AdS/CFT. This line of thought was advocated in \cite{Shuryak2007}. The general approach consists of the study of black hole formation in AdS due to the collision of shock waves. These shock waves completely stop shortly after the collision, which has been interpreted as the holographic counterpart of the nuclear stopping due to strong coupling effects \cite{Albacete2008}. The process of horizon formation in AdS due to time-dependent perturbations on the boundary, oriented to the study of far-from-equilibrium isotropization in the plasma, was recently explored by numerical methods \cite{Chesler2008b}. Moreover, by means of so-called {\it holographic renormalization}, it is possible to extract information on the real-time dynamics of the energy-momentum tensor of the gauge theory just after the impact ({\it e.g.} an estimate of the thermalization time as a function of the energy density) from the collision of shock waves \cite{Grumiller2008a}.

\subsection{Bjorken hydrodynamics}

The drop of quark-gluon plasma in the experimental setup undergoes a rapid expansion. From the field theory point of view, the condition of boost invariance in the central rapidity region, together with some other mild assumptions, leads to the specific pattern of Bjorken flow \cite{Bjorken1983}. A holographic dual of this phenomenon was unveiled in \cite{Janik2006}, where the late-time dynamics reproduces the $T(\tau) \sim \tau^{-1/3}$ cooling law. Moreover, studying the conditions that are necessary to avoid the appearance of singularities, the famous ratio $\eta/s = 1/(4\pi)$ was obtained in \cite{Janik2006b}. This analysis can be extended to higher orders in a long wavelength expansion, and this was shown to provide a powerful tool to extract all kinds of transport coefficients \cite{Bhattacharyya2008,Baier2008}.

~

There is clearly more to be discussed in this new and exciting arena, however here we simply refer the interested reader to the quickly growing library of research papers.


\begin{theacknowledgments}
JDE wishes to thank the organizers of the XIII Mexican School of Particles and Fields and specially Alberto G\"uijosa and Elena C\'aceres for inviting him to deliver his lectures in a wonderful and stimulating ambience. This work is supported in part by MICINN and FEDER (grant FPA2008-01838), by Xunta de Galicia (Conseller\'\i a de Educaci\'on and grant PGIDIT06PXIB206185PR), and by the Spanish Consolider-Ingenio 2010 Programme CPAN (CSD2007-00042). JDE is a {\it Ram\'on y Cajal} Research Fellow. JPS is supported by a {\it Juan de la Cierva} fellowship. The Centro de Estudios Cient\'\i ficos (CECS) is funded by the Chilean Government through the Millennium Science Initiative and the Centers of Excellence Base Financing Program of Conicyt. CECS is also supported by a group of private companies which at present includes Antofagasta Minerals, Arauco, Empresas CMPC, Indura, Naviera Ultragas and Telef\'onica del Sur.
\end{theacknowledgments}

\end{document}